\documentclass[aps,pra,twocolumn,10pt,superscriptaddress]{revtex4-2}
\usepackage[utf8]{inputenc}
\usepackage{amssymb}
\usepackage{amsmath}
\usepackage{xcolor}
\usepackage{graphicx}
\usepackage{dsfont}
\usepackage{physics}
\usepackage{color}
\usepackage{hyperref}
\usepackage{stmaryrd}
\usepackage{pifont}
\usepackage{multirow}
\usepackage{longtable}
\hypersetup{
    colorlinks=true,
    linkcolor=purple,
    citecolor=teal,     
    urlcolor=cyan,
    pdftitle={},
}

\begin{document}

\title{Lattice surgery-based logical state teleportation via noisy links}
\author{Áron Márton}
\affiliation{Institute for Theoretical Nanoelectronics (PGI-2), Forschungszentrum J\"{u}lich, 52428 J\"{u}lich, Germany}
\affiliation{Institute for Quantum Information, RWTH Aachen University, 52056 Aachen, Germany} 
\author{Luis Colmenarez}
\affiliation{Institute for Theoretical Nanoelectronics (PGI-2), Forschungszentrum J\"{u}lich, 52428 J\"{u}lich, Germany}
\affiliation{Institute for Quantum Information, RWTH Aachen University, 52056 Aachen, Germany}
\author{Lukas B\"{o}deker}
\affiliation{Institute for Theoretical Nanoelectronics (PGI-2), Forschungszentrum J\"{u}lich, 52428 J\"{u}lich, Germany}
\affiliation{Institute for Quantum Information, RWTH Aachen University, 52056 Aachen, Germany}
\author{Markus M\"{u}ller}
\affiliation{Institute for Theoretical Nanoelectronics (PGI-2), Forschungszentrum J\"{u}lich, 52428 J\"{u}lich, Germany}
\affiliation{Institute for Quantum Information, RWTH Aachen University, 52056 Aachen, Germany}

\begin{abstract}
    For planar architectures surface code-based quantum error correction is one of the most promising approaches to fault-tolerant quantum computation. This is partially due to the variety of fault-tolerant logical protocols that can be implemented in two dimensions using local operations. One such protocol is the lattice surgery-based logical state teleportation, which transfers a logical quantum state from an initial location on a quantum chip to a target location through a linking region of qubits. This protocol serves as a basis for higher-level routines, such as the entangling CNOT gate or magic state injection. In this work we investigate the correctability phase diagram of this protocol for distinct error rates inside the surface code patches and within the linking region. We adopt techniques from statistical physics to describe the numerically observed crossover regime between correctable and uncorrectable quantum error correction phases, where the correctability depends on the separation between the initial and target locations. We find that inside the crossover regime the correctability-threshold lines decay as a power law with increasing separation, which we explain accurately using a finite-size scaling analysis. Our results indicate that the logical state teleportation protocol can tolerate much higher noise rates in the linking region compared to the bulk of the surface code patches, provided the separation between the positions is relatively small.
\end{abstract}

\maketitle

\section{Introduction}

Quantum computers hold the promise to efficiently solve several problems that are intractable for classical computers. Building such quantum devices that operate reliably in the presence of unavoidable noise requires the implementation of quantum error correction (QEC). For platforms where qubits are restricted to a two-dimensional grid with only nearest-neighbour connectivity the surface code \cite{Kitaev_2003,bravyi1998quantumcodeslatticeboundary,Fowler_2012,Dennis_2002} is the leading candidate for QEC due to its high threshold, scalability, and planar connectivity. In recent years, promising experiments have realized QEC with the surface code \cite{acharya2024quantumerrorcorrectionsurface,google2023suppressing,eickbusch2024demonstratingdynamicsurfacecodes,Marques_2021,krinner2022realizing,Het_nyi_2024} using superconducting qubits and also fault-tolerant logical computation has been demonstrated based on other QEC codes using neutral atoms \cite{Bluvstein_2023,rodriguez2024experimentaldemonstrationlogicalmagic} and trapped ions \cite{C_Ryan_Anderson_2024}. 

Another advantage of the surface code for planar architectures is the ability to realize a universal fault-tolerant logical gate set in two spatial dimensions. Single-qubit Clifford gates can be implemented by braiding the corners of surface code patches \cite{Brown_2017,Gidney_2024,Geh_r_2024}, while multi-qubit Clifford operations are realized through multi-qubit Pauli measurements. These Pauli measurements are naturally available between neighbouring surface code patches via lattice surgery \cite{Horsman_2012,Chamberland_2022,Chamberland_2022_v2,Litinski_2019}. To achieve universality, fault-tolerant initialization of magic states is required, which can be accomplished using distillation protocols \cite{Bravyi_2005,Bravyi_2012,rodriguez2024experimentaldemonstrationlogicalmagic}, magic state cultivation \cite{gidney2024magicstatecultivationgrowing}, or other approaches \cite{Brown_2020,Gupta_2024,Laubscher_2019}.

\begin{figure*}[t]
    \centering
    \includegraphics[width=.8\textwidth]{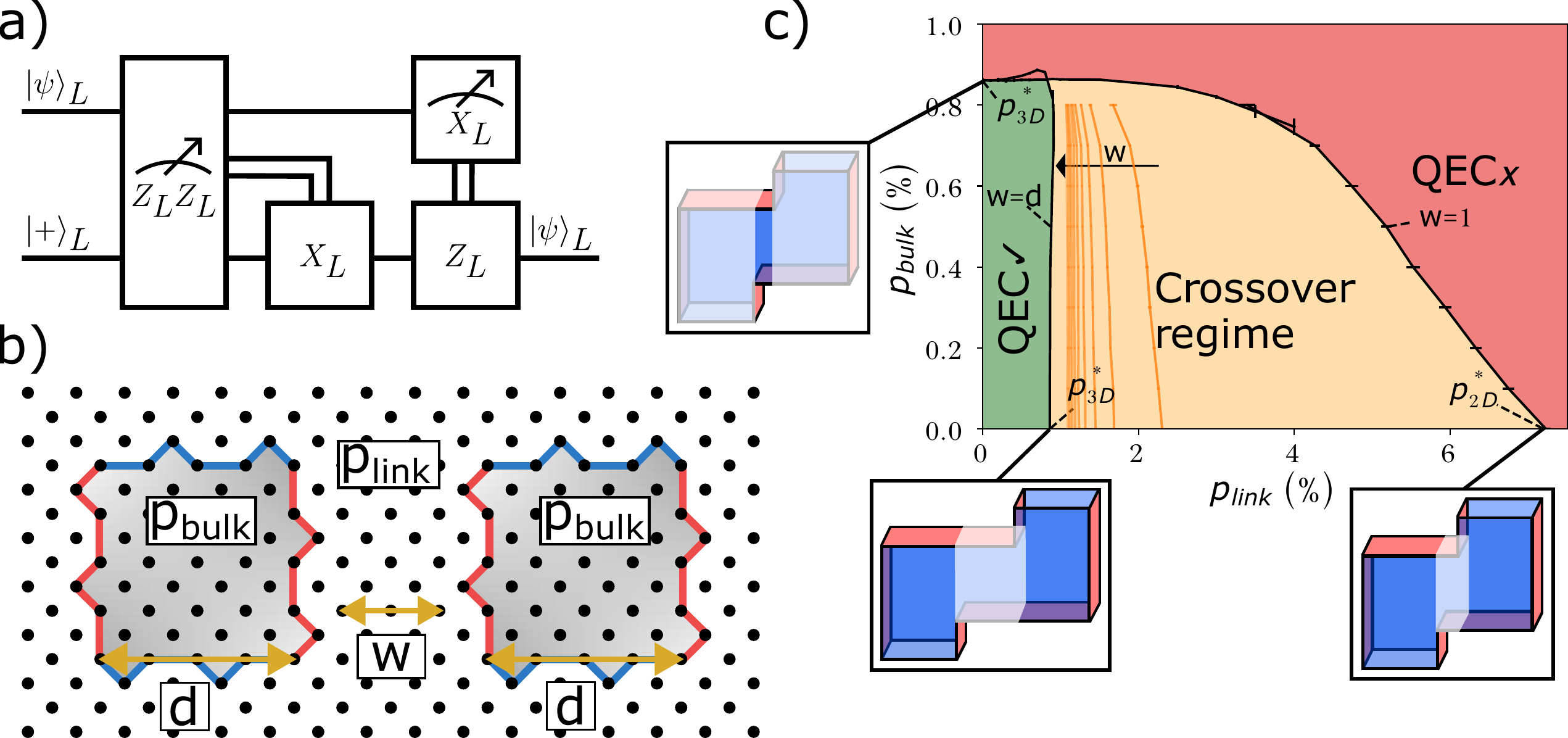}
    \caption{Illustration of the main concepts of this work. (a) The logical circuit of the measurement-based state teleportation protocol. The first logical qubit is initialized in an arbitrary $\ket{\psi}_L$ state, while the second, at the target position, is initialized in $\ket{+}_L$. A joint $Z_LZ_L$ measurement and subsequently, an $X_L$ measurement on the first logical qubit is performed. If the error-corrected outcome of the $Z_LZ_L$ measurement is $-1$ an $X_L$ correction is applied to the second logical qubit, while if the outcome of the $X_L$ measurement is $-1$ a $Z_L$ correction is applied. (b) The arrangement of surface code patches on a quantum chip. The initial distance-$d$ patch (left) and the target distance-$d$ patch (right) are separated by a linking region containing $w$ columns of data qubits. The error rate inside the bulks of the patches (grey area) is $p_{bulk}$, while in the linking region it is $p_{link}$. In the figure $d=5$ and $w=3$. (c) The numerically determined phase diagram of the lattice surgery-based teleportation of the $\ket{+_L}$ state under circuit-level noise. In the QEC\ding{51} phase scaling up the fault distance of the protocol (the code distance together with the number of measurement rounds during the lattice surgery) decreases the probability of logical errors regardless of the linking region's width. In the QEC\ding{55} phase scaling up increases the logical error rate. In the crossover regime, the effect of scaling up on the logical error rate depends on the width of the linking region. More precisely, threshold lines (orange) follow a power-law behaviour: $p_{link}^*(w,p_{bulk}) = p^*_{3D} + z(p_{bulk})w^{-1/\nu_3}$. For low $p_{link}$ rates bulk errors determine the critical behaviour. These errors are located in the three-dimensional spacetime volumes above the surface code patches. On the contrary, for low $p_{bulk}$ rates the critical behaviour is dominated by the link errors, with the corresponding spacetime segment above the linking region. For $w=1$ this segment is two dimensional, while for $w=d$ this volume is three-dimensional. This difference in the dimensionality results in an extended crossover regime, where the correctability depends on $w$. The spacetime regions where dominating errors occur are depicted as white volumes for different parameter regimes.}
    \label{fig:main}
\end{figure*}

While lattice surgery-based operations have been investigated both experimentally \cite{besedin2025realizinglatticesurgerydistancethree,lacroix2024scalinglogiccolorcode,Erhard_2021,C_Ryan_Anderson_2024,Het_nyi_2024} and numerically \cite{Chamberland_2022,Chamberland_2022_v2,Ramette2023Faulttolerant,sinclair2024faulttolerantopticalinterconnectsneutralatom}, their behavior under spatially inhomogeneous errors has remained relatively unexplored.
In this paper we investigate this scenario for a specific protocol: lattice surgery-based logical state teleportation. This protocol not only enables beyond-nearest-neighbor connectivity between logical qubits, but also serves as a basis for higher-level protocols such as the entangling logical CNOT gate or magic state injection via lattice surgery. In the lattice surgery-based logical state telportation an arbitrary logical state $\ket{\psi}_L$ is teleported from one 
location of the quantum chip to another. The logical circuit for this protocol is depicted in Fig.~\ref{fig:main}. In our setup, the initial and target locations of the protocol are separated by an intermediate linking region of width $w$. It is known that in the bulk of surface code patches, physical qubits and gates must operate below a certain threshold \cite{Fowler_2012,Dennis_2002,Wang_2003} to achieve a correctable QEC regime, in which the logical error rate can be arbitrarily suppressed by increasing the code size. However, in the linking region, the error rates of qubits and gates can exceed the bulk threshold while maintaining correctability, as has been shown for lattice surgery with a single line of noisy communication links \cite{Ramette2023Faulttolerant}.

We extend this result for an arbitrarily broad linking region by numerically determining the phase diagram of the logical teleportation protocol for varying error rates in the bulk and linking region. We show that if the error rates are below the bulk threshold $p_{3D}^*$ the teleportation protocol is in the correctable (QEC\ding{51}) phase, where increasing the code distance suppresses logical errors arbitrarily, regardless of the separation between logical qubits. Conversely, when bulk error rates exceed $p_{3D}^*$, the protocol immediately transitions to the uncorrectable (QEC\ding{55}) phase, where increasing the code distance makes the performance worse. 

Interestingly, for bulk error rates below $p_{3D}^*$, a crossover regime emerges between QEC\ding{51} and QEC\ding{55} phases, where correctability depends on the separation between the surface code patches, $w$. The emergence of this regime can be understood by analyzing the 2+1 dimensional spacetime diagram representing the lattice surgery-based logical state teleportation protocol (see Fig.~\ref{fig:main}). For low $p_{link}$ error rates, the critical behaviour is determined by the bulk errors, which occur in three-dimensional spacetime volumes above the surface code patches. In contrast, for low $p_{bulk}$ rates, the link errors drive the transition; these errors occur on a two-dimensional surface for $w=1$, and in a three-dimensional volume for $w=d$. This difference in the dimensionality leads to distinct critical error rates, with $p^*_{2D} > p^*_{3D}$. For finite, but constant values of $w$, the dominant errors occur in a quasi-two-dimensional spacetime volume, resulting in $w$-dependent threshold lines following a power-law decay as $w\rightarrow \infty$ (see Fig.~\ref{fig:main}). We adopt techniques from statistical physics to describe the crossover regime, and apply a finite-size scaling analysis to derive this power-law decay of the threshold lines.

We extend the results of \cite{Ramette2023Faulttolerant} and show that the teleportation protocol tolerates higher noise in the linking region than in the bulk, even for logical qubit separations larger than $w=1$. We also show that the decaying threshold lines follow a power-law as the separation between logical qubits increases. These results may relax design constraints for quantum chips that host multiple logical qubits, as it permits noisier gates and physical qubits in the linking regions.

The rest of the paper is structured as follows: Sec.~\ref{sec:surface_code_intro} introduces the surface code and describes the steps of the lattice surgery-based logical state teleportation protocol. Sec.~\ref{sec:phase_diagram} qualitatively explains the structure of the phase diagram of the teleportation protocol and explores the implications of the dimensionality of different spacetime regions. Finally, Sec.~\ref{sec:crossover} examines how the threshold depends on the separation of logical qubits in the crossover regime and outlines the procedure used to extract the phase diagram from the numerical data.

\section{Surface codes and logical state teleportation} \label{sec:surface_code_intro}

\begin{figure}
    \centering
    \includegraphics[width=0.45\textwidth]{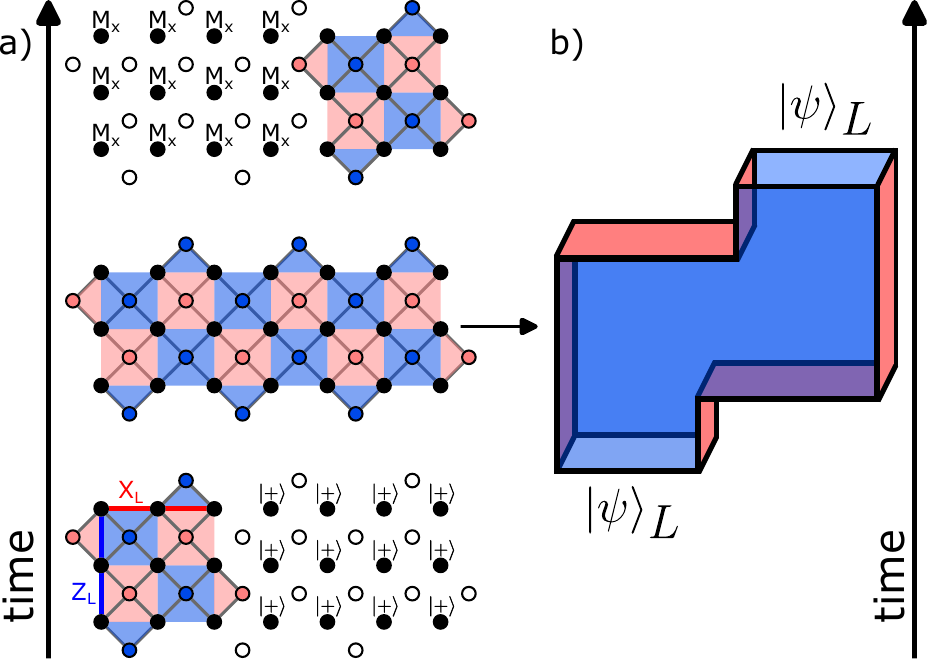}
    \caption{The three steps and the spacetime diagrammatic representation of the lattice surgery-based logical state teleportation protocol. (a) A surface code patch with $X$-stabilizers (red plaquettes) and $Z$-stabilizers (blue plaquettes) is initialized in an arbitrary logical state $\ket{\psi_L}$ on the left part of the chip. The logical operators are depicted as red ($X_L$) and blue ($Z_L$) strings. The data qubits in the linking region and on the right are initialized in $\ket{+}$. In the next step all the colored stabilizers are measured $d$ times. Finally, the data qubits in the linking region and on the left are measured in the $X$ basis. In this example, $d=3$ and $w=1$. (b) The spacetime diagram representation of the protocol. Vertical surfaces on the left and right correspond to $X$-type boundaries (red), while the front and back correspond to $Z$-type boundaries, as only $X$-stabilizers are present on the left and right, and $Z$-stabilizers on the front and back. The horizontal surfaces are also $X$-type boundaries, because only $X$-stabilizer values can be constructed from initializing/measuring in the $X$-basis.}
    \label{fig:surgery}
\end{figure}

In a distance-$d$ rotated surface code patch \cite{Bombin_2007} a single logical qubit is encoded into $d^2$ physical data qubits. The $d^2-1$ stabilizers of the code are measured with the help of $d^2-1$ auxiliary qubits through a syndrome extraction circuit \cite{Dennis_2002,Fowler_2012}. A rotated surface code patch is visualized in Fig.~\ref{fig:surgery}. The logical $X_L$ and $Z_L$ operators correspond to the product of single-qubit $X$ and $Z$ operators, respectively, acting on the top row and left column of data qubits, as it is depicted in Fig.~\ref{fig:surgery}. It is important to note that the product of a logical operator and stabilizers is also a logical operator, e.g. logical $X_L$ ($Z_L$) can also be located on the bottom (right) side of the code.

To teleport a logical qubit encoded in a surface code patch from one location of the quantum chip to another, we consider a measurement-based circuit as shown in Fig.~\ref{fig:main}. A surface code patch at the target location is first initialized in the $\ket{+}_L$ state. Next, a joint $Z_LZ_L$ measurement is performed via lattice surgery \cite{Horsman_2012}, and finally the logical qubit at the initial position is measured in the $X_L$ basis. Up to Pauli corrections this circuit teleports arbitrary logical state $\ket{\psi}_L$ from the initial to the target location.

The joint $Z_LZ_L$ measurement is performed as follows:
\begin{enumerate}
    \item The qubits in the linking region that assume the role of data qubits in the extended rectangular patch during the surgery, are initialized in the $\ket{+}^{\otimes d\cdot w}$ state.
    \item $d$ rounds of X- and Z-stabilizer measurements are performed, covering all stabilizers defined within the rectangular patch that includes both the original patches and the linking region.
    \item The qubits in the linking region that assume the role of data qubits in the extended rectangular patch during the surgery, are measured in the $X$-basis.
\end{enumerate}
To determine the outcome of the logical measurement, the $Z$-stabilizer measurement outcomes in the linking region must be multiplied, yielding a result that can still be corrected by the decoder \cite{Horsman_2012}. This procedure relies on the fact that the product of the $Z$-stabilizers in the linking region is exactly the joint $Z_LZ_L$ operator. Multiple rounds of stabilizer measurements are required to protect the logical measurement outcome against errors that corrupt the readout of the stabilizers. $d$ rounds ensure the fault distance to be $d$ for the whole protocol \cite{Bomb_n_2023,Domokos_2024,Chamberland_2022}.

In the measurement-based logical state teleportation protocol (see Fig.~\ref{fig:main}), the surface code patch at the target position is initialized in the $\ket{+}_L$ state and the initial patch is measured in the $X$-basis, irrespective of the teleported state $\ket{\psi}_L$. Therefore, the initialization and final logical $X_L$-measurement can be integrated into the lattice surgery process without loss of generality. Instead of initializing the $\ket{+}_L$ state, the data qubits of the target patch are initialized in the $\ket{+}^{\otimes d^2}$ state. This approach is valid because the $Z_LZ_L$ measurement commutes with the stabilizer measurements and after the lattice surgery the stabilizers of this patch are measured anyway. Similarly, rather than measuring the stabilizers of the initial patch after the surgery, the data qubits are directly measured in the $X$-basis, with the $X$-stabilizer and $X_L$ operator values reconstructed from these measurements. Our simplified procedure is illustrated in Fig.~\ref{fig:surgery}.

Some interesting properties of surface code-based logical protocols can be visualized with spacetime diagrams \cite{Bomb_n_2023,Gidney_2024,Domokos_2024}. These diagrams depict the time evolution of the constituent surface codes by highlighting the spacetime locations of different boundary types. At $X$-type boundaries only $X$ stabilizers are present, while at $Z$-type boundaries only $Z$ ones. 
For our purposes the spacetime diagram representation of the lattice surgery-based state teleportation protocol (see Fig.~\ref{fig:surgery}) is important, because the dimensionality of the spacetime regions where different errors can occur can be visualized.

\section{The structure of the correctability phase diagram} \label{sec:phase_diagram}

We numerically determine the phase diagram of the teleportation protocol, considering different error rates in the linking region and inside the bulks of the surface code patches (see Fig.~\ref{fig:main}). 
This section provides a qualitative explanation of the phase diagram's structure and its relation to the dimensionality of various regions in the spacetime diagram of the protocol.

We numerically simulate the logical state teleportation protocol at the circuit level using STIM \cite{Gidney2021stimfaststabilizer}. In our error model each gate is followed by a single- or two-qubit depolarizing channel, and qubit initialization and measurements may also fail. All three error mechanisms occur with the same probability, $p_{bulk}$ within the surface code patches and $p_{link}$ in the linking region. Specifically, we simulate the fault-tolerant teleportation of the logical $\ket{+}_L$ state. Here the protocol is preceded by the fault-tolerant initialization of $\ket{+}_L$ on the left patch. This is realized by the initialization of the $\ket{+}^{\otimes d^2}$ state followed by a stabilizer measurement round. Similarly, at the end of the protocol a fault-tolerant measurement is performed in the $X_L$ basis on the teleported state. This is realized by a stabilizer measurement round after the surgery and a final measurement of the data qubits of the right patch in the $X$ basis. In Appendix~\ref{appendix:zero_state} we discuss the teleportation of the logical $\ket{0_L}$ state in detail. We use MWPM \cite{Dennis_2002,edmonds_1965,higgott2022pymatching,higgott2023sparse} to decode the syndrome produced by the stabilizer measurements of the protocol. Our STIM circuits and all the data are publicly available at \cite{muller_2025_15257403}.

When one of the error rates, either $p_{bulk}$ or $p_{link}$, is fixed, the transition from the correctable to the uncorrectable regime can be characterized by a threshold value in the respective other, unfixed error parameter. This statement holds only if the fixed error rate is sufficiently small to ensure that the success probability for the logical teleportation undergoes a phase transition. For low $p_{link}$ rates, the errors in the bulk will dominate the transition, and these errors are located in the spacetime regions above the surface code patches. Because the number of measurement rounds is scaling together with the code distance, the size of these spacetime volumes is approximately $d\times d\times d$. Conversely, for low $p_{bulk}$ rates the transition is dominated by the errors in the linking region. These errors are located in the spacetime region above the linking region with the size $w\times d\times d$. For the smallest possible $w=1$ this is a two-dimensional surface, while for $w$ scaled together with $d$ this is a three dimensional volume. The relevant spacetime regions for different parameter regimes are shown in Fig.~\ref{fig:main}.

Examining the dimensionality of the relevant spacetime regions provides key insights into the structure of the phase diagram. The threshold value depends on the dimensionality of the associated spacetime region: for a two-dimensional surface, the threshold $p^*_{2D}$ is much higher than the threshold $p^*_{3D}$ for a three-dimensional volume \cite{Wang_2003,Ramette2023Faulttolerant,tan2024resiliencesurfacecodeerror}. For example under phenomenological noise with MWPM decoding $p^*_{3D} \approx 2.9\%$ and $p^*_{2D} \approx 10.3\%$ \cite{Wang_2003}. This dimensional analysis explains the sharp transition from the QEC\ding{51} phase to the QEC\ding{55} phase for small $p_{link}$, since there the relevant spacetime region is consistently three-dimensional, regardless of $w$.  It also clarifies the boundaries of the crossover region: when $w=d$ the link error threshold is the three-dimensional threshold, $p^*_{3D}$, while for $w=1$ it corresponds to the two-dimensional threshold $p^*_{2D}$. The connection between the threshold value and the dimensionality of the relevant spacetime region is summarized in Table~\ref{tab:threshold_dimensions}. For constant $w > 1$ the link error threshold falls between $p^*_{3D}$ and $p^*_{2D}$. The precise dependence on $w$ is detailed in Sec.~\ref{sec:crossover}.
\begin{table}[!h]
    \centering
    \begin{tabular}{|p{2cm}|p{2cm}|p{2.4cm}|p{1.8cm}|}
    \hline
        The fixed error parameter & Linking region's width  & Relevant spacetime region & Threshold \\
    \hline
        \multirow{2}{*}{$p_{link} \ll p^*_{3D}$} & $w = 1$ & $3D$ & $p^*_{3D}$ \\
        & $w = d$ & $3D$ & $p^*_{3D}$ \\
        \hline
        \multirow{2}{*}{$p_{bulk} \ll p^*_{3D}$} & $w = 1$ & $2D$ & $p^*_{2D}$ \\
        & $w = d$ & $3D$ & $p^*_{3D}$ \\
    \hline
    \end{tabular}
    \caption{The dimensionality of the relevant spacetime region and the corresponding threshold value for different parameter regimes. For low $p_{link}$ rates the relevant spacetime region is three-dimensional, regardless of $w$. Conversely, for low $p_{bulk}$ rates the relevant region can be two-dimensional when $w=1$.}
    \label{tab:threshold_dimensions}
\end{table}

When both link and bulk errors approach their respective threshold values, the entire spacetime diagram becomes relevant. This leads to a slight decrease in the threshold values, resulting in curvature of the phase boundaries, as reported in Ref.~\cite{Ramette2023Faulttolerant}.

\section{Finite size scaling and the crossover regime} \label{sec:crossover}

Correctability transition of QEC codes can, in many cases, be related to phase transitions of classical disordered statistical mechanical models \cite{Dennis_2002,Wang_2003,Chubb_2021,Bombin_2012,Katzgraber_2009,lyons2024understandingstabilizercodeslocal,Vodola2022fundamental,rispler2024randomcoupledplaquettegaugemodel}. This connection enables the application of techniques from statistical physics to investigate the correctability phase diagram. In this section, we employ a finite-size scaling analysis to characterize the behavior of threshold lines within the crossover regime, a method that has already been used in the context of QEC \cite{sriram2024nonuniformnoiseratesgriffiths}.

For finite system sizes the failure rate of the logical state teleportation protocol not only depends on strength of the underlying physical errors, but also on the geometrical parameters of the spacetime diagram, $d$ and $w$. Under the assumptions of finite-size scaling, which is well-motivated by the underlying statistical mechanical mapping, the logical error rate collapses into a single-variable scaling function, where the scaling variable incorporates the geometrical parameters and the physical error rate. Using this universal collapse we derive the $w$ dependence of the shifting threshold in the crossover regime. Furthermore, we detail the procedure used to extract the threshold values from the numerical data.
\begin{figure*}[t]
    \centering
    \includegraphics[width=0.85\textwidth]{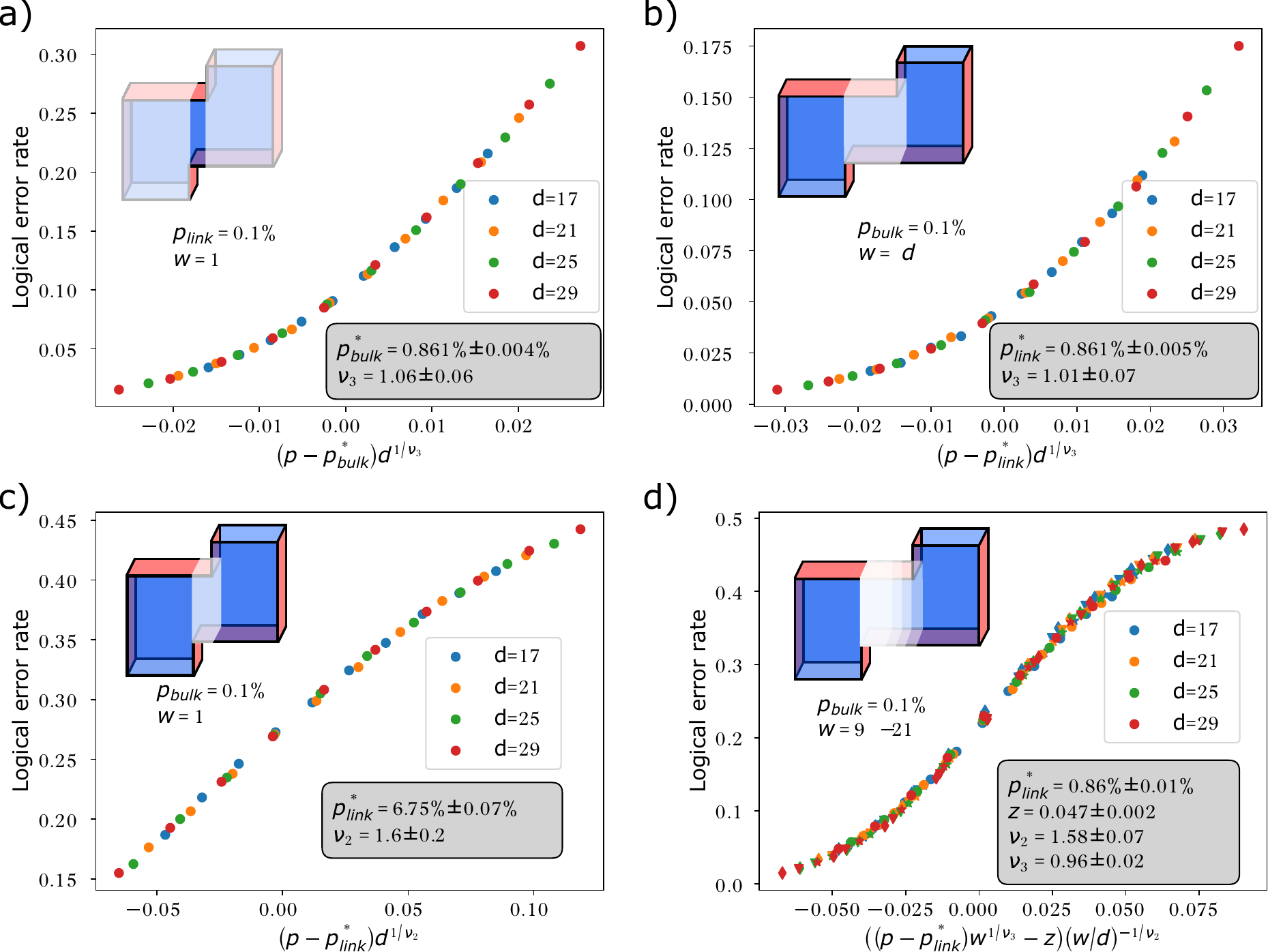}
    \caption{The universal collapse of the logical error rate. In a) for a fixed $p_{link} =0.1\%$ and for $w=1$ we determine the bulk noise threshold and $\nu_3$ using the three-dimensional scaling function, Eq.~\eqref{eq:3D}. In b) we determine the link noise threshold and $\nu_3$ for $p_{bulk}=0.1\%$ and $w=d$ using Eq.~\eqref{eq:3D}. In c) we show the link noise threshold and $\nu_3$ for $p_{bulk}=0.1\%$ and $w=1$ using Eq.~\eqref{eq:2d}. In d) we display the critical parameters of the crossover regime for $p_{bulk}=0.1\%$ using Eq.~\eqref{eq:2.5d}. The data collapse is shown for code distances $d=17,21,25,29$, in all sub-plots. In d) we analyze the different widths $w=9(\text{\ding{108}}),13(\text{\ding{72}}),17(\blacktriangledown),21(\blacklozenge)$ for each distance.}
    \label{fig:collapse}
\end{figure*}

For the sake of completeness we first outline the finite-size scaling of a three-dimensional volume with linear size $d$ and physical error rate $p$ \cite{Wang_2003}. The results of this analysis apply when the relevant region of the spacetime diagram is a three-dimensional volume, where errors occur with probability $p$. The key assumption of finite-size scaling is that for sufficiently large system sizes, the logical error rate depends only on the fraction of length scales:
\begin{align}
    p_L(p,d) = f(d/\xi_{3D}).
\end{align}
Here, $\xi_{3D}$ represents the system's correlation length, which diverges near the threshold as:
\begin{align}
    \xi_{3D} \sim (p-p^*_{3D})^{-\nu_3}.
\end{align}
This leads to the scaling form of the logical error rate in a $d\times d\times d$ region:
\begin{align} \label{eq:3D}
    p_L(p,d) = \Phi_{3D}\big((p-p^*_{3D})d^{1/\nu_3}\big).
\end{align}

Similarly, the scaling form of the logical error rate in a two-dimensional surface with linear size $d$, and physical error rate $p$ is given by:
\begin{align} \label{eq:2d}
    p_L(p,d) = \Phi_{2D}\big((p-p^*_{2D})d^{1/\nu_2}\big),
\end{align}
where the universal exponent $\nu_2$ and the non-universal threshold differ from the three-dimensional case.

To describe a quasi-two-dimensional slab of size $w\times d\times d$, the additional length scale, $w$, must be incorporated. In this case, the finite size scaling assumption becomes:
\begin{align} \label{eq:fs_assumption_2}
    p_L(p,d,w) = f(w/\xi_{3D}, w/d).
\end{align}
In the limit of $w/d \rightarrow 0$, the logical error rate should recover the two-dimensional scaling form, Eq.~\eqref{eq:2d}:
\begin{align} \label{eq:w_const_assumption}
    p_L(p,d,w)_{w/d \rightarrow 0} = \Phi_w\big((p-p^*(w))d^{1/\nu_2}\big),
\end{align}
where the non-universal quantities such as the threshold and the scaling function's form can depend on $w$. Starting from Eq.~\eqref{eq:fs_assumption_2} and Eq.~\eqref{eq:w_const_assumption} the scaling form of the logical error rate in the limit of $w/d \rightarrow 0$ can be derived \cite{sriram2024nonuniformnoiseratesgriffiths} (see the derivation in Appendix~\ref{appendix:scaling_derivation}), and it is given as:
\begin{align} \label{eq:2.5d}
    p_L(p,d,w) = \Phi\Big(\big((p-p^*_{3D})w^{1/\nu_3}-z\big)(w/d)^{-1/\nu_2}\Big).\,
\end{align}
where $z$ is a non-universal parameter of the scaling variable. By comparing Eq.~\eqref{eq:2.5d} with Eq.~\eqref{eq:w_const_assumption} the $w$ dependence of the shifting threshold is obtained as:
\begin{align} \label{eq:shifting_threshold}
    p^*(w)= p^*_{3D} + zw^{-1/\nu_3}.
\end{align}

We determine the critical parameters by collapsing the numerical data into the appropriate scaling form of the logical error rate. We identify the optimal parameters by minimizing an objective function that quantifies the quality of the data collapse. Details of this procedure are outlined in \cite{Zabalo_2020} and in Appendix~\ref{appendix:collapse}. Depending on the dimensionality of the relevant region in the spacetime diagram (Table~\ref{tab:threshold_dimensions}), we apply the corresponding scaling function Eq.~\eqref{eq:3D}, Eq.~\eqref{eq:2d}, or Eq.~\eqref{eq:2.5d} for 3D, 2D and quasi-2D, respectively. Examples of the resulting collapsed data for different parameter regimes are shown in Fig.~\ref{fig:collapse}.
\begin{figure}
    \centering
    \includegraphics[width=0.48\textwidth]{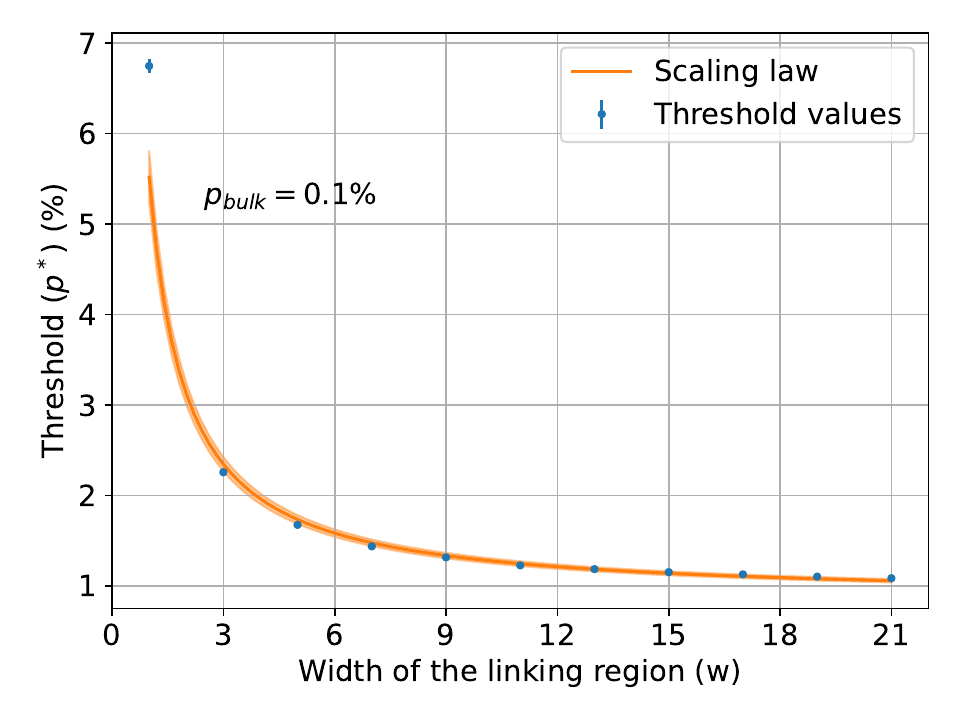}
    \caption{The $w$-dependence of the threshold of the $\ket{+_L}$ state teleportation in the crossover regime for a fixed $p_{bulk}=0.1\%$ error rate. The continuous orange curve shows the crossover scaling law, described by Eq.~\eqref{eq:shifting_threshold}, with critical parameters $p^*_{3D} =0.86(1)\%$, $z=0.047(2)$, $\nu_2=1.58(7)$ and $\nu_3=0.96(2)$. The region shaded in light orange color indicates the associated error bars. The blue dots represent threshold values determined independently for each width value $w$.}
    \label{fig:shifting}
\end{figure}

To show the $w$-dependence of the shifting threshold in the crossover regime we plot Eq.~\eqref{eq:shifting_threshold} for a fixed $p_{bulk}$ rate in Fig.~\ref{fig:shifting}. The critical parameters ($p^*_{3D}$, $z$ and $\nu_3$) are determined by collapsing the data as shown in Fig.~\ref{fig:collapse}. Additionally, in Fig.~\ref{fig:shifting} we display the threshold values determined independently for each $w$ using Eq.~\eqref{eq:w_const_assumption} as the scaling function. 

Figure~\ref{fig:shifting} demonstrates that the scaling law described by Eq.~\eqref{eq:shifting_threshold} accurately captures the $w$-dependence of the shifting threshold for large $w$. However, for $w=1$ there is a significant difference between the true threshold value and the one predicted by the scaling law. We suspect that this discrepancy arises because, for small $w$, the finite size scaling assumption described by Eq.~\eqref{eq:fs_assumption_2} breaks down. The true threshold value, $p^*_{2D}$, is actually higher than the value predicted by the crossover scaling law, $p^*_{3D} +z$ (at least for the $\ket{+_L}$ state teleportation; results for $\ket{0_L}$ can be found in Appendix~\ref{appendix:zero_state}.). Overall, our results highlight, from a practical point of view, the robustness of the state teleportation protocol against errors in the linking region. Furthermore, from a methodological perspective, they underline the suitabality of the chosen statistical physics analysis to quantitatively describe the threshold behaviour and interplay of bulk and linking region error rates.

\section{Conclusion}

In this work, we have determined the correctability phase diagram of a lattice surgery-based logical state teleportation protocol. Here, we have analyzed the effect and interplay of distinct physical error rates inside the bulk of the surface code patches and in the linking region. Our analysis shows that for low error probability in the linking region a sharp transition occurs between the correctable QEC\ding{51} and the uncorrectable QEC\ding{55} phases, in which logical teleportation reliably succeeds or fails, respectively. However, for low bulk error rates the QEC\ding{51} and the QEC\ding{55} phases are separated by an extended crossover regime, where the overall correctability of the protocol depends on the separation $w$ between logical qubits. Using finite size scaling arguments, we have found that the shifting threshold within the crossover regime is well-described by the following crossover scaling law ansatz, $p^*(w)=p^*_{3D} + zw^{-1/\nu_3}$.

These findings suggest that the lattice surgery-based logical state teleportation protocol tolerates significantly higher noise rates in the linking region compared to the bulk when the separation between the initial and target patches is small. This result may allow to relax quality requirements for quantum processors with multiple logical qubits, enabling the use of noisier qubits and gates in linking regions without compromising correctability. 

For large separations, this drastic increase in the threshold vanishes. An open question is whether the lattice surgery protocol can be modified to maintain a high threshold in the linking region even for large separations. It will also valuable to extend the presented analysis for the study of other, potentially more complex, logical qubit operations needed for scalable universal fault-tolerant quantum computation, such as lattice-surgery-based CNOT gate operations, or logical T-gate injection protocols by means of lattice surgery approaches.

STIM circuits and all the data are publicly available at \cite{muller_2025_15257403}.

\section{Acknowledgements}

We gratefully acknowledge support by the European Union’s Horizon Europe research and innovation program under Grant Agreement Number 101114305 (``MILLENION-SGA1'' EU Project), the US Army Research Office through Grant Number W911NF-21-1-0007, the Office of the Director of National Intelligence (ODNI), Intelligence Advanced Research Projects Activity (IARPA) and the Army Research Office, under the Entangled Logical Qubits program through Cooperative Agreement Number W911NF-
23-2-0212. The views and conclusions contained in
this document are those of the authors and should not
be interpreted as representing the official policies, either
expressed or implied, of IARPA, the Army Research Office, or the U.S. Government. The U.S. Government is
authorized to reproduce and distribute reprints for Government purposes notwithstanding any copyright notation
herein. This research is also part of the Munich Quantum Valley (K-8), which is supported by the Bavarian state government with funds from the Hightech Agenda Bayern Plus. We also acknowledge support by the Deutsche Forschungsgemeinschaft (DFG, German Research Foundation) under Germany’s Excellence Strategy ‘Cluster of Excellence Matter and Light for Quantum Computing (ML4Q) EXC 2004/1’ 390534769, and by the ERC Starting Grant QNets through Grant No.~804247. Numerical computations have been performed using the PGI-2 compute cluster at Forschungszentrum J\"ulich.

\bibliographystyle{apsrev4-2-custom}
\bibliography{main}

\appendix

\renewcommand{\arraystretch}{0.5}

\section{Teleportation of the logical $\ket{0_L}$ state} \label{appendix:zero_state}

\begin{figure}[!h]
    \centering
    \includegraphics[width=0.9\linewidth]{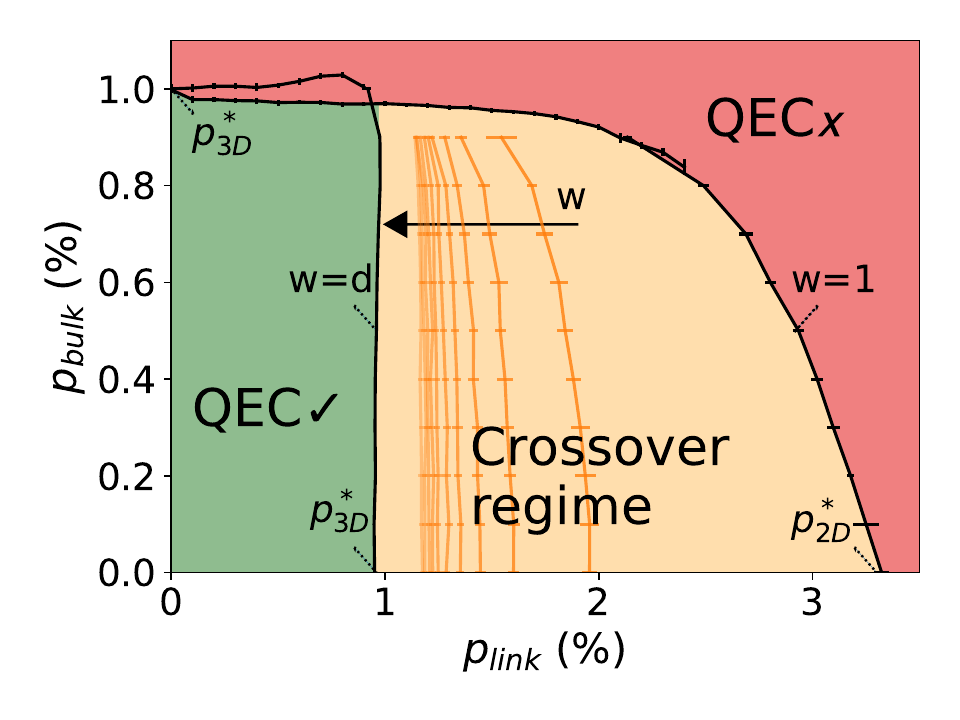}
    \caption{The numerically determined phase diagram of the lattice surgery-based teleportation of the logical $\ket{0_L}$ state.}
    \label{fig:phase_diagram_zero}
\end{figure}

\begin{figure}
    \centering
    \includegraphics[width=0.9\linewidth]{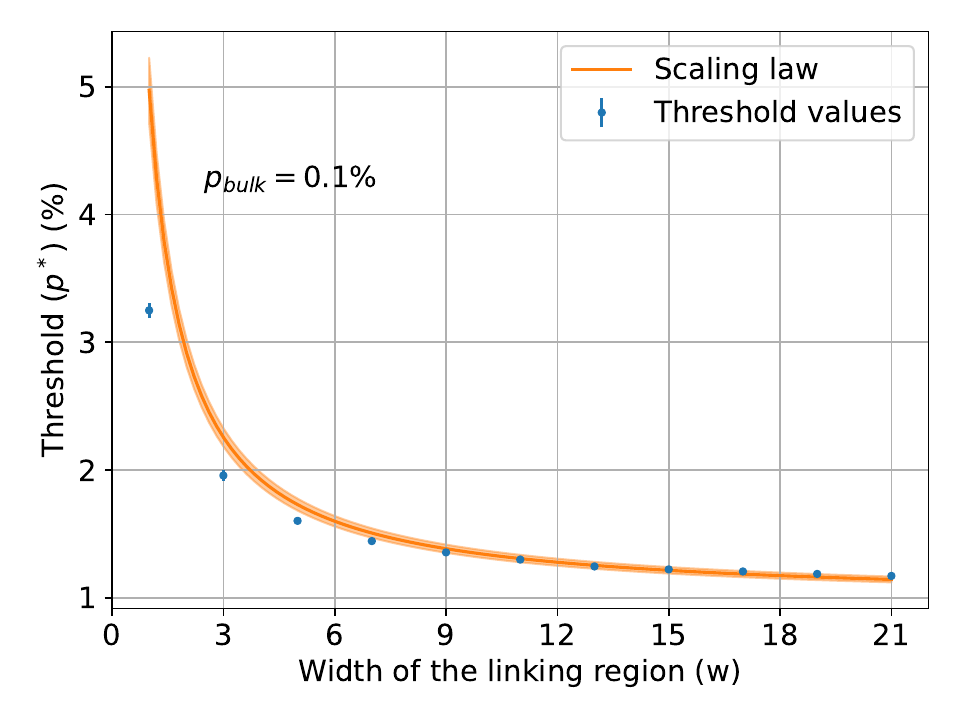}
    \caption{The $w$-dependence of the threshold of the $\ket{0_L}$ state teleportation in the crossover regime for a fixed $p_{bulk}=0.1\%$ error rate. The continuous orange curve shows the crossover scaling law, described by Eq.~\eqref{eq:shifting_threshold}, with critical parameters $p^*_{3D} =0.97(2)\%$, $z=0.040(1)$, $\nu_2=1.57(4)$ and $\nu_3=0.97(2)$. The region shaded in light orange color indicates the associated error bars. The blue dots represent threshold values determined independently for each $w$.}
    \label{fig:crossocer_scaling_zero}
\end{figure}

In the main text we discuss our results only for the $\ket{+_L}$ state teleportation. Here we show the phase diagram and the $w$-dependence of the shifting threshold for the $\ket{0_L}$ state in Figs.~\ref{fig:phase_diagram_zero} and \ref{fig:crossocer_scaling_zero}, respectively. While the qualitative features of the two phase diagrams are similar, we highlight some minor differences that may be of interest.

The exact locations of phase boundaries differ, because the underlying protocol is not symmetric under exchanging $X$ and $Z$. The most noticeable difference is in the numerical values of $p^*_{2D}$ ($p^*_{2D} \approx 7.2\%$ for $\ket{+_L}$ and $p^*_{2D} \approx 3.3\%$ for $\ket{0_L}$). We believe this discrepancy arises because the logical error strings responsible for the failure of $\ket{+_L}$ state teleportation are spacelike, whereas those causing the failure of $\ket{0_L}$ state teleportation are timelike above the linking region. Therefore, the difference between the timelike and spacelike error rates results in a much higher $p^*_{2D}$ for the $\ket{+_L}$ state teleportation.

As opposed to $\ket{+_L}$ state teleportation, the crossover scaling law overestimates the true phase boundary, as shown in Fig.~\ref{fig:crossocer_scaling_zero} for $p_{bulk} = 0.1\%$. Interestingly, the orange curves in Figs.~\ref{fig:crossocer_scaling_zero} and \ref{fig:shifting} predicts the $\ket{0}_L$ and $\ket{+_L}$ phase boundaries to be close to each other, however, the true values differ significantly. This observation shows that the difference between timelike and spacelike error rates has more drastic consequences for $w=1$ than for larger $w$-s.

We also observe a ``bump" in the $w=d$ curve in Fig.~\ref{fig:phase_diagram_zero}, so the phase boundary for $w=d$ is above the phase boundary for $w=1$, for low, but finite $p_{link}$ rates. This effect can also be observed in the phase diagram of the $\ket{+_L}$ state teleportation protocol, however there the gap between the phase boundaries is much smaller. We believe this is a finite-size effect, as the true threshold in the thermodynamic limit should not increase with increasing either of the physical error rates. %Nevertheless, this interesting effect warrants further investigation.

\section{Derivation of the scaling variable in the crossover regime}\label{appendix:scaling_derivation}

To derive Eq.~\eqref{eq:2.5d} from Eqs.~\eqref{eq:fs_assumption_2} and~\eqref{eq:w_const_assumption}, we follow the approach of \cite{sriram2024nonuniformnoiseratesgriffiths}. We begin by rewriting Eq.~\eqref{eq:fs_assumption_2} in an equivalent form:
\begin{align} \label{eq:scaling_assump}
    p_L(p,d,w) = \Phi\big((p-p^*_{3D})w^{1/\nu_3},w/d\big).
\end{align}
Taking $w/d = 0$, the logical error rate simplifies to a function of a single scaling variable:
\begin{align} \label{eq:wd=0}
    p_L(p,d,w)_{w/d = 0} = \Phi'\big((p-p^*_{3D})w^{1/\nu_3}\big).
\end{align}
However, in the limit $w/d \rightarrow 0$ the logical error rate can also be expressed in the form of Eq.~\eqref{eq:w_const_assumption}:
\begin{align} \label{eq:wd-0}
    p_L(p,d,w)_{w/d \rightarrow 0} = \Phi_w\big((p-p^*(w))d^{1/\nu_2}\big).
\end{align}
From Eq.~\eqref{eq:wd-0} we know that $\Phi'\big((p-p^*_{3D})w^{1/\nu_3}\big)$ must be singular at $p=p^*(w)$. Denoting the singular point as:
\begin{align}
    z = (p^*(w)-p^*_{3D})w^{1/\nu_3},
\end{align}
we can express the shifting threshold as:
\begin{align} \label{eq:shifting}
    p^*(w) = p^*_{3D} + zw^{-1/\nu_3}.
\end{align}

To determine the scaling form of the logical error rate in the limit $w/d \rightarrow 0$, we first analyze the dependence of $\Phi_w\big((p-p^*(w))d^{1/\nu_2}\big)$ on $w$, then express $p^*(w)$ using Eq.~\eqref{eq:shifting}. Consistency with Eq.~\eqref{eq:scaling_assump} requires that $\Phi_w\big((p-p^*(w))d^{1/\nu_2}\big)$ depends on $w$ only through a multiplicative factor:
\begin{align}
    \Phi_w\big((p-p^*(w))d^{1/\nu_2}\big) = \Phi\big(w^\alpha(p-p^*(w))d^{1/\nu_2}\big),
\end{align}
where $\alpha$ is to be determined. Substituting Eq.~\eqref{eq:shifting} the scaling variable can be expressed as:
\begin{align}
    w^\alpha(p-p^*(w))d^{1/\nu_2} = w^\alpha(p-p^*_{3D}-zw^{-1/\nu_3})d^{1/\nu_2}.
\end{align}
This expression can only be consistent with Eq.~\eqref{eq:scaling_assump} if $\alpha = 1/\nu_3 - 1/\nu_2$. Herewith, we have derived Eq.~\eqref{eq:2.5d}, the scaling form of the logical error rate in the limit $w/d \rightarrow 0$. 

\section{Data collapse} \label{appendix:collapse}

To determine the critical parameters by collapsing the data, we follow a procedure outlined in \cite{Zabalo_2020}. In the investigated cases, the logical error rate depends only on a scaling variable, which is determined by a set of critical parameters, $\underline{c}$. Consequently, the logical error rate can be expressed as:
\begin{align}
    p_L(x(\underline{c})).
\end{align}
The numerical data consists of logical error rates, $p_L^i$, with corresponding error bars, $\sigma_i$, for each set of $\{p_{link}/p_{bulk},d,w\}$. For a fixed set $\underline{c}$, we calculate the scaling variable $x_i$ for each set of $\{p_{link}/p_{bulk},d,w\}$ and order the data such that
\begin{align}
    x_{i-1} \leq x_i \leq x_{i+1}. 
\end{align}
To determine the optimal critical parameters that yield the best data collapse, we minimize the following objective function:
\begin{align}
    O(\underline{c}) = \dfrac{1}{n-2}\sum_{j=2}^{n-1}\Big(\dfrac{p_L^i-\bar{p}_L^i}{\Delta(p_L^i-\bar{p}_L^i)}\Big)^2,
\end{align}
where $\bar{p}_L^i$ and $\Delta(p_L^i-\bar{p}_L^i)$ are defined as:
\begin{align}
    &\bar{p}_L^i = \dfrac{(x_{i+1}-x_{i})p_L^{i-1} + (x_{i}-x_{i-1})p_L^{i+1}}{x_{i+1}-x_{i-1}} \\
    &\Delta(p_L^i-\bar{p}_L^i) = \sigma_i^2 + \Big(\dfrac{x_{i+1}-x_i}{x_{i+1}-x_{i-1}}\sigma_{i-1}\Big)^2 \\ \nonumber
    &+ \Big(\dfrac{x_{i}-x_{i-1}}{x_{i+1}-x_{i-1}}\sigma_{i+1}\Big)^2.
\end{align}
This procedure is equivalent to minimizing the deviation (weighted by the variance) of each point $(x_i,p_L^i)$ from the line determined by its adjacent points $(x_{i-1},p_L^{i-1})$ and $(x_{i+1},p_L^{i+1})$.

To estimate the uncertainty in the critical parameters we use a bootstrapping approach. We regenerate the dataset 100 times, assuming that the number of logical failures follows a binomial distribution with mean $N\cdot p_L^i$, where $N$ is the number of shots. For each regenerated dataset, we determine a corresponding set of critical parameters. The uncertainty in each parameter is then estimated as three times the standard deviation across these 100 trials.

\section{Numerical data}

We summarize the coordinates and the corresponding critical exponents of the numerically determined points of the threshold lines shown in Figs.~\ref{fig:main} and \ref{fig:phase_diagram_zero} in Tables~\ref{tab:data_plus} and \ref{tab:data_zero}. Moreover, we summarize the critical parameters of the crossover regimes for different $p_{bulk}$ error rates in Tables~\ref{tab:collapse_plus} and \ref{tab:collapse_zero}.
\begin{table*}
\centering
    \begin{tabular}{|c|cccccccccc|}
        \hline 
        &&&&&$w=1$&&&&& \\
        \hline 
        $p_{link}$ (\%) & 7.23(6) & 6.75(7) & 6.33(8) & 5.94(8) & 5.52(9) & 5.18(7) & 4.74(8) & 4.27(7) & 3.4(1) & \\
        $p_{bulk}$ (\%) & 0 & 0.1 & 0.2 & 0.3 & 0.4 & 0.5 & 0.6 & 0.7 & 0.8 & \\
        $\nu$ & 1.7(1) & 1.6(2) & 1.5(1) & 1.5(3) & 1.6(2) & 1.6(1) & 1.7(2) & 1.5(1) & 1.5(1) & \\
        \hline
        $p_{link}$ (\%) & 0 & 0.1 & 0.2 & 0.3 & 0.4 & 0.5 & 0.6 & 0.7 & 0.8 & 0.9 \\
        $p_{bulk}$ (\%) & 0.862(4) & 0.861(4) & 0.861(3) & 0.860(4) & 0.861(4) & 0.862(3) & 0.862(3) & 0.863(3) & 0.863(3) & 0.864(4) \\
        $\nu$ & 1.09(6) & 1.06(6) & 1.05(6) & 1.05(5) & 1.08(6) & 1.06(6) & 1.04(5) & 1.05(5) & 1.02(5) & 1.04(5) \\
        \hline
        $p_{link}$ (\%) & 1 & 1.5 & 2 & 2.5 & 3 & 3.5 & 4 & & & \\
        $p_{bulk}$ (\%) & 0.863(3) & 0.863(3) & 0.854(3) & 0.844(4) & 0.820(5) & 0.78(2) & 0.75(2) & & & \\
        $\nu$ & 1.03(5) & 1.04(5) & 1.03(5) & 1.07(6) & 1.13(8) & 1.3(2) & 1.3(3) & & & \\
        \hline 
        &&&&&$w=3$&&&&& \\
        \hline
        $p_{link}$ (\%) & 2.30(2) & 2.26(2) & 2.20(2) & 2.15(2) & 2.11(2) & 2.05(2) & 1.99(2) & 1.89(2) & 1.67(5) & \\
        $p_{bulk}$ (\%) & 0 & 0.1 & 0.2 & 0.3 & 0.4 & 0.5 & 0.6 & 0.7 & 0.8 & \\
        $\nu$ & 1.5(2) & 1.5(2) & 1.5(2) & 1.5(2) & 1.4(2) & 1.5(2) & 1.4(2) & 1.4(1) & 1.6(3) & \\
        \hline 
        &&&&&$w=5$&&&&& \\
        \hline
        $p_{link}$ (\%) & 1.69(1) & 1.67(1) & 1.64(1) & 1.63(1) & 1.60(2) & 1.58(1) & 1.55(1) & 1.51(1) & 1.38(2) & \\
        $p_{bulk}$ (\%) & 0 & 0.1 & 0.2 & 0.3 & 0.4 & 0.5 & 0.6 & 0.7 & 0.8 & \\
        $\nu$ & 1.5(1) & 1.5(1) & 1.5(1) & 1.5(1) & 1.6(2) & 1.5(1) & 1.5(1) & 1.4(1) & 1.4(2) & \\
        \hline 
        &&&&&$w=7$&&&&& \\
        \hline
        $p_{link}$ (\%) & 1.45(1) & 1.44(1) & 1.42(1) & 1.413(9) & 1.41(1) & 1.39(1) & 1.37(1) & 1.35(1) & 1.26(2) & \\
        $p_{bulk}$ (\%) & 0 & 0.1 & 0.2 & 0.3 & 0.4 & 0.5 & 0.6 & 0.7 & 0.8 & \\
        $\nu$ & 1.55(9) & 1.6(1) & 1.6(1) & 1.5(1) & 1.47(9) & 1.5(1) & 1.49(9) & 1.5(1) & 1.5(1) & \\
        \hline 
        &&&&&$w=9$&&&&& \\
        \hline
        $p_{link}$ (\%) & 1.318(9) & 1.315(8) & 1.303(9) & 1.29(1) & 1.295(9) & 1.28(1) & 1.28(1) & 1.26(1) & 1.19(1) & \\
        $p_{bulk}$ (\%) & 0 & 0.1 & 0.2 & 0.3 & 0.4 & 0.5 & 0.6 & 0.7 & 0.8 & \\
        $\nu$ & 1.56(9) & 1.52(8) & 1.46(8) & 1.52(8) & 1.52(8) & 1.49(8) & 1.45(9) & 1.36(9) & 1.4(1) & \\
        \hline 
        &&&&&$w=11$&&&&& \\
        \hline
        $p_{link}$ (\%) & 1.238(8) & 1.226(8) & 1.229(8) & 1.226(8) & 1.23(1) & 1.22(1) & 1.214(8) & 1.206(9) & 1.15(1) & \\
        $p_{bulk}$ (\%) & 0 & 0.1 & 0.2 & 0.3 & 0.4 & 0.5 & 0.6 & 0.7 & 0.8 & \\
        $\nu$ & 1.60(9) & 1.52(8) & 1.50(8) & 1.49(7) & 1.6(1) & 1.49(8) & 1.42(8) & 1.40(6) & 1.43(8) & \\
        \hline 
        &&&&&$w=13$&&&&& \\
        \hline
        $p_{link}$ (\%) & 1.189(9) & 1.185(9) & 1.180(9) & 1.173(8) & 1.175(9) & 1.172(8) & 1.172(9) & 1.17(1) & 1.131(8) & \\
        $p_{bulk}$ (\%) & 0 & 0.1 & 0.2 & 0.3 & 0.4 & 0.5 & 0.6 & 0.7 & 0.8 & \\
        $\nu$ & 1.52(9) & 1.49(8) & 1.55(9) & 1.48(8) & 1.52(8) & 1.41(8) & 1.47(8) & 1.40(9) & 1.40(8) & \\
        \hline 
        &&&&&$w=15$&&&&& \\
        \hline
        $p_{link}$ (\%) & 1.15(1) & 1.151(8) & 1.144(8) & 1.146(8) & 1.144(8) & 1.140(9) & 1.139(8) & 1.133(8) & 1.10(1) & \\
        $p_{bulk}$ (\%) & 0 & 0.1 & 0.2 & 0.3 & 0.4 & 0.5 & 0.6 & 0.7 & 0.8 & \\
        $\nu$ & 1.49(9) & 1.51(7) & 1.54(9) & 1.50(7) & 1.50(7) & 1.48(9) & 1.46(7) & 1.32(6) & 1.35(9) & \\
        \hline 
        &&&&&$w=17$&&&&& \\
        \hline
        $p_{link}$ (\%) & 1.127(9) & 1.127(9) & 1.123(9) & 1.121(7) & 1.117(8) & 1.119(8) & 1.119(8) & 1.115(9) & 1.09(1) & \\
        $p_{bulk}$ (\%) & 0 & 0.1 & 0.2 & 0.3 & 0.4 & 0.5 & 0.6 & 0.7 & 0.8 & \\
        $\nu$ & 1.54(8) & 1.59(9) & 1.54(9) & 1.47(6) & 1.50(8) & 1.50(7) & 1.42(7) & 1.35(7) & 1.4(1) & \\
        \hline 
        &&&&&$w=19$&&&&& \\
        \hline
        $p_{link}$ (\%) & 1.11(1) & 1.102(7) & 1.101(7) & 1.105(9) & 1.100(8) & 1.096(9) & 1.100(8) & 1.100(7) & 1.073(9) & \\
        $p_{bulk}$ (\%) & 0 & 0.1 & 0.2 & 0.3 & 0.4 & 0.5 & 0.6 & 0.7 & 0.8 & \\
        $\nu$ & 1.5(1) & 1.50(7) & 1.48(7) & 1.53(7) & 1.5(1) & 1.47(8) & 1.42(7) & 1.32(6) & 1.29(8) & \\
        \hline 
        &&&&&$w=21$&&&&& \\
        \hline
        $p_{link}$ (\%) & 1.094(7) & 1.085(8) & 1.09(1) & 1.086(9) & 1.09(1) & 1.083(9) & 1.078(9) & 1.086(9) & 1.072(9) & \\
        $p_{bulk}$ (\%) & 0 & 0.1 & 0.2 & 0.3 & 0.4 & 0.5 & 0.6 & 0.7 & 0.8 & \\
        $\nu$ & 1.54(8) & 1.47(8) & 1.56(9) & 1.50(9) & 1.5(1) & 1.50(8) & 1.36(6) & 1.29(6) & 1.32(7) & \\
        \hline
        &&&&&$w=d$&&&&& \\
        \hline
        $p_{link}$ (\%) & 0.862(6) & 0.861(5) & 0.864(5) & 0.864(4) & 0.868(6) & 0.881(4) & 0.893(4) & 0.911(4) & 0.907(8) & 0.75(6) \\
        $p_{bulk}$ (\%) & 0 & 0.1 & 0.2 & 0.3 & 0.4 & 0.5 & 0.6 & 0.7 & 0.8 & 0.9 \\
        $\nu$ & 1.03(9) & 1.01(7) & 1.02(8) & 1.04(7) & 1.03(7) & 0.99(8) & 0.89(6) & 0.79(6) & 0.77(7) & 1.5(6) \\
        \hline
        $p_{link}$ (\%) & 0 & 0.1 & 0.2 & 0.3 & 0.4 & 0.5 & 0.6 & 0.7 & 0.8 & 0.9 \\
        $p_{bulk}$ (\%) & 0.861(4) & 0.861(4) & 0.862(5) & 0.865(3) & 0.868(3) & 0.872(3) & 0.879(3) & 0.887(3) & 0.881(3) & 0.82(2)  \\
        $\nu$ & 1.06(6) & 1.12(7) & 1.05(7) & 1.09(6) & 1.06(6) & 1.05(6) & 0.98(5) & 0.95(5) & 0.93(6) & 1.3(4) \\
        \hline  
        \end{tabular}
     \caption{The numerically determined coordinates ($p_{link}$,$p_{bulk}$) and the corresponding critical exponents of the threshold lines for different $w$-s for the teleportation of the $\ket{+_L}$ state.}
     \label{tab:data_plus}
\end{table*}

\begin{table*}
\centering
    \begin{tabular}{|c|cccccccccccc|}
        \hline 
        &&&&&&$w=1$&&&&&& \\
        \hline 
        $p_{link}$ (\%) & 3.32(4) & 3.25(6) & 3.18(2) & 3.10(3) & 3.02(3) & 2.93(3) & 2.80(2) & 2.69(3) & 2.49(3) & 2.13(2) & 0 & 0.1 \\
        $p_{bulk}$ (\%) & 0 & 0.1 & 0.2 & 0.3 & 0.4 & 0.5 & 0.6 & 0.7 & 0.8 & 0.9 & 0.998(7) & 0.978(7) \\
        $\nu$ & 1.8(4) & 1.4(6) & 1.6(1) & 1.6(1) & 1.6(1) & 1.6(1) & 1.6(2) & 1.5(1) & 1.6(1) & 1.7(2) & 1.2(2) & 1.1(1) \\
        \hline
        $p_{link}$ (\%) & 0.2 & 0.3 & 0.4 & 0.5 & 0.6 & 0.7 & 0.8 & 0.9 & 1 & 1.1 & 1.2 & 1.3 \\
        $p_{bulk}$ (\%) & 0.978(6) & 0.976(6) & 0.975(7) & 0.972(6) & 0.972(4) & 0.972(5) & 0.969(5) & 0.969(5) & 0.967(4) & 0.969(5) & 0.966(5) & 0.962(5) \\
        $\nu$ & 1.0(1) & 1.05(8) & 1.0(1) & 1.1(1) & 1.02(7) & 0.97(8) & 1.04(8) & 0.95(7) & 0.95(8) & 0.98(7) & 0.94(4) & 0.96(4) \\
        \hline
        $p_{link}$ (\%) & 1.4 & 1.5 & 1.6 & 1.7 & 1.8 & 1.9 & 2 & 2.1 & 2.2 & 2.3 & 2.4 & \\
        $p_{bulk}$ (\%) & 0.961(5) & 0.956(5) & 0.953(4) & 0.949(5) & 0.942(6) & 0.932(6) & 0.921(7) & 0.90(1) & 0.883(8) & 0.869(8) & 0.84(2) &  \\
        $\nu$ & 0.93(4) & 0.93(5) & 0.95(3) & 0.94(4) & 0.97(5) & 0.98(5) & 0.99(6) & 1.06(9) & 1.04(7) & 1.1(1) & 1.1(1) & \\
        \hline
        &&&&&&$w=3$&&&&&& \\
        \hline
        $p_{link}$ (\%) & 1.96(4) & 1.96(4) & 1.94(5) & 1.91(4) & 1.88(3) & 1.84(4) & 1.81(4) & 1.75(4) & 1.69(2) & 1.54(7) & & \\
        $p_{bulk}$ (\%) & 0 & 0.1 & 0.2 & 0.3 & 0.4 & 0.5 & 0.6 & 0.7 & 0.8 & 0.9 &  &  \\
        $\nu$ & 1.5(2) & 1.7(2) & 1.7(3) & 1.7(3) & 1.6(2) & 1.6(3) & 1.6(3) & 1.5(2) & 1.6(2) & 1.9(5) & & \\
        \hline
        &&&&&&$w=5$&&&&&& \\
        \hline
        $p_{link}$ (\%) & 1.60(3) & 1.60(3) & 1.59(3) & 1.57(3) & 1.56(4) & 1.54(02) & 1.53(4) & 1.49(3) & 1.46(3) & 1.36(3) & & \\
        $p_{bulk}$ (\%) & 0 & 0.1 & 0.2 & 0.3 & 0.4 & 0.5 & 0.6 & 0.7 & 0.8 & 0.9 &  &  \\
        $\nu$ & 1.6(2) & 1.7(2) & 1.6(2) & 1.6(2) & 1.7(3) & 1.5(2) & 1.7(3) & 1.5(2) & 1.5(2) & 1.6(3) & & \\
        \hline
        &&&&&&$w=7$&&&&&& \\
        \hline
        $p_{link}$ (\%) & 1.45(3) & 1.44(2) & 1.43(3) & 1.43(2) & 1.41(3) & 1.41(2) & 1.39(3) & 1.37(3) & 1.34(2) & 1.28(3) & & \\
        $p_{bulk}$ (\%) & 0 & 0.1 & 0.2 & 0.3 & 0.4 & 0.5 & 0.6 & 0.7 & 0.8 & 0.9 &  &  \\
        $\nu$ & 1.6(2) & 1.8(2) & 1.6(2) & 1.6(2) & 1.6(2) & 1.7(2) & 1.5(2) & 1.5(2) & 1.4(2) & 1.6(2) & & \\
        \hline
        &&&&&&$w=9$&&&&&& \\
        \hline
        $p_{link}$ (\%) & 1.35(2) & 1.35(2) & 1.34(2) & 1.34(3) & 1.34(2) & 1.33(2) & 1.32(2) & 1.30(2) & 1.28(1) & 1.22(1) & & \\
        $p_{bulk}$ (\%) & 0 & 0.1 & 0.2 & 0.3 & 0.4 & 0.5 & 0.6 & 0.7 & 0.8 & 0.9 &  &  \\
        $\nu$ & 1.6(1) & 1.7(2) & 1.6(2) & 1.6(3) & 1.6(1) & 1.6(1) & 1.6(1) & 1.5(1) & 1.4(1) & 1.4(1) & & \\
        \hline
        &&&&&&$w=11$&&&&&& \\
        \hline
        $p_{link}$ (\%) & 1.29(2) & 1.30(2) & 1.29(2) & 1.29(2) & 1.28(2) & 1.27(2) & 1.26(2) & 1.25(1) & 1.25(2) & 1.20(2) & & \\
        $p_{bulk}$ (\%) & 0 & 0.1 & 0.2 & 0.3 & 0.4 & 0.5 & 0.6 & 0.7 & 0.8 & 0.9 &  &  \\
        $\nu$ & 1.6(1) & 1.7(2) & 1.6(2) & 1.6(1) & 1.6(1) & 1.5(1) & 1.5(1) & 1.4(1) & 1.4(1) & 1.5(2) & & \\
        \hline
        &&&&&&$w=13$&&&&&& \\
        \hline
        $p_{link}$ (\%) & 1.25(2) & 1.25(1) & 1.25(2) & 1.24(1) & 1.24(2) & 1.24(2) & 1.23(1) & 1.23(2) & 1.22(1) & 1.18(1) & & \\
        $p_{bulk}$ (\%) & 0 & 0.1 & 0.2 & 0.3 & 0.4 & 0.5 & 0.6 & 0.7 & 0.8 & 0.9 &  &  \\
        $\nu$ & 1.5(2) & 1.5(1) & 1.6(1) & 1.5(1) & 1.5(1) & 1.5(2) & 1.5(1) & 1.5(2) & 1.4(1) & 1.42(9) & & \\
        \hline
        &&&&&&$w=15$&&&&&& \\
        \hline
        $p_{link}$ (\%) & 1.22(2) & 1.22(2) & 1.23(2) & 1.22(2) & 1.21(1) & 1.22(02) & 1.21(01) & 1.20(1) & 1.19(1) & 1.16(1) & & \\
        $p_{bulk}$ (\%) & 0 & 0.1 & 0.2 & 0.3 & 0.4 & 0.5 & 0.6 & 0.7 & 0.8 & 0.9 &  &  \\
        $\nu$ & 1.5(1) & 1.6(2) & 1.6(1) & 1.5(1) & 1.5(1) & 1.6(1) & 1.51(9) & 1.4(1) & 1.3(1) & 1.4(1) & & \\
        \hline
        &&&&&&$w=17$&&&&&& \\
        \hline
        $p_{link}$ (\%) & 1.21(2) & 1.21(1) & 1.20(2) & 1.20(2) & 1.20(2) & 1.20(2) & 1.19(1) & 1.19(1) & 1.18(1) & 1.15(1) & & \\
        $p_{bulk}$ (\%) & 0 & 0.1 & 0.2 & 0.3 & 0.4 & 0.5 & 0.6 & 0.7 & 0.8 & 0.9 &  &  \\
        $\nu$ & 1.6(2) & 1.6(1) & 1.5(1) & 1.5(1) & 1.5(1) & 1.6(2) & 1.5(1) & 1.5(1) & 1.3(1) & 1.3(2) & & \\
        \hline
        &&&&&&$w=19$&&&&&& \\
        \hline
        $p_{link}$ (\%) & 1.19(1) & 1.19(1) & 1.19(1) & 1.19(2) & 1.19(2) & 1.18(1) & 1.17(1) & 1.17(1) & 1.17(1) & 1.15(1) & & \\
        $p_{bulk}$ (\%) & 0 & 0.1 & 0.2 & 0.3 & 0.4 & 0.5 & 0.6 & 0.7 & 0.8 & 0.9 &  &  \\
        $\nu$ & 1.5(1) & 1.6(1) & 1.5(1) & 1.6(1) & 1.6(2) & 1.5(1) & 1.48(9) & 1.4(1) & 1.31(9) & 1.4(1) & & \\
        \hline 
        &&&&&&$w=21$&&&&&& \\
        \hline
        $p_{link}$ (\%) & 1.18(1) & 1.17(1) & 1.17(2) & 1.17(1) & 1.16(1) & 1.17(1) & 1.17(1) & 1.16(1) & 1.16(1) & 1.14(1) & & \\
        $p_{bulk}$ (\%) & 0 & 0.1 & 0.2 & 0.3 & 0.4 & 0.5 & 0.6 & 0.7 & 0.8 & 0.9 &  &  \\
        $\nu$ & 1.6(1) & 1.5(1) & 1.5(1) & 1.5(1) & 1.5(1) & 1.5(1) & 1.5(1) & 1.42(9) & 1.32(8) & 1.4(1) & & \\
        \hline 
        &&&&&&$w=d$&&&&&& \\
        \hline
        $p_{link}$ (\%) & 0.951(5) & 0.949(6) & 0.955(6) & 0.953(6) & 0.954(7) & 0.960(5) & 0.963(5) & 0.969(4) & 0.977(6) & 0.976(7) &  & \\
        $p_{bulk}$ (\%) & 0 & 0.1 & 0.2 & 0.3 & 0.4 & 0.5 & 0.6 & 0.7 & 0.8 & 0.9 &  &  \\
        $\nu$ & 1.03(7) & 1.1(1) & 1.00(9) & 1.1(1) & 1.1(1) & 0.99(9) & 1.00(7) & 0.98(8) & 0.93(9) & 1.0(1) & & \\
        \hline
        $p_{link}$ (\%) & 0 & 0.1 & 0.2 & 0.3 & 0.4 & 0.5 & 0.6 & 0.7 & 0.8 & 0.9 &  & \\
        $p_{bulk}$ (\%) & 1.000(9) & 1.002(8) & 1.006(6) & 1.006(7) & 1.003(8) & 1.008(5) & 1.016(7) & 1.026(6) & 1.029(6) & 1.004(6) & &  \\
        $\nu$ & 1.1(2) & 1.1(2) & 1.1(1) & 1.1(2) & 1.1(2) & 1.0(1) & 1.0(1) & 0.94(09) & 0.84(09) & 0.9(1) & & \\
        \hline  
        \end{tabular}
     \caption{The numerically determined coordinates ($p_{link}$,$p_{bulk}$) and the corresponding critical exponents of the threshold lines for different $w$-s for the teleportation of the $\ket{0_L}$ state.}
     \label{tab:data_zero}
\end{table*}

\begin{table*}
    \centering
    \begin{tabular}{|c|ccccccccc|}
    \hline
    $p_{bulk}$ (\%) & 0 & 0.1 & 0.2 & 0.3 & 0.4 & 0.5 & 0.6 & 0.7 & 0.8  \\
    $p^*_{3D}$ (\%) & 0.87(2) & 0.86(1) & 0.87(2) & 0.88(1) & 0.89(2) & 0.890(5) & 0.91(1) & 0.93(1) & 0.94(1)  \\
    $z$ & 0.046(3) & 0.047(2) & 0.046(2) & 0.047(2) & 0.046(2) & 0.046(1) & 0.046(1) & 0.046(3) & 0.045(3) \\
    $\nu_2$ & 1.6(1) & 1.58(7) & 1.58(4) & 1.59(3) & 1.56(4) & 1.57(2) & 1.58(3) & 1.54(6) & 1.6(3) \\
    $\nu_3$ & 0.98(3) & 0.96(2) & 0.96(2) & 0.94(2) & 0.93(3) & 0.93(1) & 0.91(1) & 0.89(3) & 0.86(4)  \\
    \hline
    \end{tabular}
    \caption{The numerically determined critical parameters of the data collapse inside the crossover regime with $w=9,11,13,15,17,19$, and $21$, for the teleportation of the $\ket{+_L}$ state.}
    \label{tab:collapse_plus}
\end{table*}

\begin{table*}
    \centering
    \begin{tabular}{|c|cccccccccc|}
    \hline
    $p_{bulk}$ (\%) & 0 & 0.1 & 0.2 & 0.3 & 0.4 & 0.5 & 0.6 & 0.7 & 0.8 & 0.9 \\
    $p^*_{3D}$ (\%) & 0.961(9) & 0.97(2) & 0.97(2) & 0.96(2) & 0.98(1) & 0.982(6) & 0.99(2) & 0.997(8) & 1.00(2) & 0.99(1) \\
    $z$ & 0.039(2) & 0.040(1) & 0.040(1) & 0.037(3) & 0.040(3) & 0.040(1) & 0.040(5) & 0.039(1) & 0.037(5) & 0.040(2) \\
    $\nu_2$ & 1.59(5) & 1.57(4) & 1.55(5) & 1.61(7) & 1.58(9) & 1.55(4) & 1.56(6) & 1.61(4) & 1.62(9) & 1.60(8) \\
    $\nu_3$ & 0.99(3) & 0.97(2) & 0.97(2) & 1.01(4) & 0.96(5) & 0.95(2) & 0.93(7) & 0.94(2) & 0.94(5) & 0.87(2) \\
    \hline
    \end{tabular}
    \caption{The numerically determined critical parameters of the data collapse inside the crossover regime with $w=9,11,13,15,17,19$, and $21$,  for the teleportation of the $\ket{0_L}$ state.}
    \label{tab:collapse_zero}
\end{table*}

\end{document}